\documentclass[twocolumn,pre,showpacs,preprintnumbers,amsmath,amssymb]{revtex4-1}

\usepackage{graphicx}
\usepackage{dcolumn}
\usepackage{bm}
\usepackage{hyperref}
\usepackage{color}

\begin{document}


\title{Mixed mode oscillation suppression states in coupled oscillators} 

\author{Debarati Ghosh}
\email{debarati.physics.2014@gmail.com}
\affiliation{Department of Physics, University of Burdwan, Burdwan 713 104, West Bengal, India.}
\author{Tanmoy Banerjee}
\thanks{Corresponding author}
\email{tbanerjee@phys.buruniv.ac.in}
\affiliation{Department of Physics, University of Burdwan, Burdwan 713 104, West Bengal, India.}

\date{\today}

\begin{abstract}
We report a new collective dynamical state, namely the {\it mixed mode oscillation suppression} state where different set of variables of a system of coupled oscillators show different types of oscillation suppression states. We identify two variants of it: The first one is a {\it mixed mode death} (MMD) state where a set of variables of a system of coupled oscillators show an oscillation death (OD) state, while the rest are in an amplitude death (AD) state under the identical parametric condition. In the second mixed death state (we refer it as the MNAD state) a nontrivial bistable AD and a monostable AD state appear simultaneously to different set of variables of a same system. We find these states in paradigmatic chaotic Lorenz system and Lorenz-like system under generic coupling schemes. We identify that while the reflection symmetry breaking is responsible for the MNAD state, the breaking of both the reflection and translational symmetries result in the MMD state. Using a rigorous bifurcation analysis we establish the occurrence of the MMD and MNAD states, and map their transition routes in parameter space. Moreover, we report the {\it first} experimental observation of the MMD and MNAD states that supports our theoretical results. We believe that this study will broaden our understanding of oscillation suppression states, subsequently, it may have applications in many real physical systems, such as laser and geomagnetic systems, whose mathematical models mimic the Lorenz system.
\end{abstract}


\maketitle 

\section{Introduction}
\label{sec:intro}
The suppression of oscillation is an important emergent behavior shown by coupled oscillators, and it has been extensively studied in a variety of fields such as physics, biology, chemistry and engineering \cite{kosprep}. Amplitude death (AD) and oscillation death (OD) are the two distinct oscillation quenching processes. In the AD state, oscillations cease and the coupled oscillators reside in a common stable steady state; thus, it induces a stable homogeneous steady state (HSS) \cite{adrev}, which was unstable otherwise. On the other hand, OD is a much more complex and completely distinct phenomena than AD \cite{kosprep}. In OD, oscillators occupy different coupling dependent steady states and thus give rise to a stable inhomogeneous steady state (IHSS). AD is a widely studied topic as a control mechanism to suppress undesirable  oscillations in laser application \cite{laser}, neuronal systems \cite{bard}, electronic circuits \cite{tanchaosad}, etc. On the other hand, OD  has strong connections and influence in the biological system, e.g., a synthetic genetic oscillator \cite{kosepl, qstr2}, a coupled-oscillator system that represents cardiovascular phenomena \cite{Vargasepl} and cellular differentiation \cite{cell}.

In the context of AD and OD the work by \citet{kosprl} deserves a special mention, because it shows that AD and OD differ in manifestation and origin. Most significantly, it reports a Turing type bifurcation that marks a {\it direct} continuous transition from AD to OD. The observations of Ref.~\cite{kosprl} have later been verified in different systems and coupling schemes and a recent burst of publications explore many aspects of the AD, OD and AD--OD transitions (see for example Refs.~\cite{kospre,*scholl4,*dana1,*kurthpre}, Refs.~\cite{tanpre1,*tanpre2}, and Ref.~\cite{tanpre3}). But, a continuous endeavor to find various aspects of AD and OD in  networks (e.g., chimera death \cite{scholl,*tanCD}), new systems (e.g., ecological system \cite{bandutta}), and new coupling schemes (e.g., amplitude dependent coupling \cite{kurthamp}) indicates the necessity and urgency of further research in this field.     

In the previous studies on oscillation suppression, all the variables of a coupled oscillators under study show either OD or AD, separately, for a certain parametric condition.  Under no condition it happens that different variables of a same system show different oscillation suppression states 
\footnote{For example, in the case of paradigmatic Stuart-Landau oscillator, for a certain parameter value, both of its variables $x$ and $y$ ($Z=x+iy$) show either OD or AD: Under no coupling condition it happens that the two variables show two different oscillation suppression states for any parameter value.}.
This is owing to the fact that either the inverse Hopf bifurcation (that leads to AD) or a symmetry breaking bifurcation (that leads to OD)  occur to all the variables of a system at a time.

At this point we ask the following open question: \textbf{\it Is it possible that different variables of a system (of coupled oscillators) show different types of oscillation suppression states under an identical parametric condition}? In this paper we indeed identify this type of {\em mixed mode oscillation suppression states} in the paradigmatic Lorenz system and Lorenz-like system under several generic coupling configurations, which were earlier studied in the context of AD and OD.

In the present paper we report two {\it mixed mode oscillation suppression states}: (i) The {\it mixed mode death} (MMD) state, where, under a similar parametric condition, a set of variables of a coupled oscillator system show OD, whereas the rest of the variables of the same coupled system show the AD state. In the chaotic Lorenz system under two different coupling schemes, namely the direct-indirect \cite{resmiad,tanpre3} and mean-field coupling \cite{tanchaosad}, \cite{shino,*st,*de}, we show that, while two of the variables (say, $x$ and $y$) show the OD state, the rest (i.e., the $z$ variable) shows an AD state. Thus, unlike the AD or OD state, this MMD state is a {\it variable selective mixture} of stable IHSS (i.e., OD) and stable HSS (i.e, AD). 
(ii) In the second mixed death state, a nontrivial bistable AD (NAD) state (occurs in $x$ and $y$ variables) and a monostable AD state (occurs in the $z$ variable) appear simultaneously--we refer this state as the MNAD state (i.e., mixture of the NAD and AD state). 

We identify that the symmetry of the system plays an important role behind the birth of the MMD and MNAD states: While the {\it reflection symmetry breaking} is responsible for the MNAD state, the {\it breaking of both reflection and translational symmetries} are responsible for the MMD state. Through a rigorous bifurcation analysis, we establish the occurrence of the MMD and MNAD states and several transition routes associated with them. To establish the generality of our results we verify all the results in a Lorenz-like system, namely the chaotic Chen system \cite{chen}. {\it To the best of our knowledge, existence of the MMD and MNAD states and the corresponding transitions associated with them have not been observed earlier for any other system or coupling configuration}. Finally, we support our results through an electronic experiment that provides the {\it first} experimental evidence of the MMD and MNAD states.

The rest of the paper is organized in the following manner: The next section considers the identical Lorenz systems under two different coupling schemes. Rigorous bifurcation analysis establish the occurrence of MMD and MNAD, and related transitions. Section \ref{sec3} reports the occurrence of mixed death states in Chen system, which is a Lorenz-like system. Experimental observation of MMD and MNAD is reported in Sec.~\ref{sec4}. Finally, Sec.~\ref{sec:con} concludes the outcome and importance of the whole study. 

\section{Coupled Lorenz systems}
\label{sec2}
\subsection{Identical Lorenz systems interacting through direct-indirect coupling}
\label{sec2a}
\noindent At first, we consider two identical chaotic Lorenz oscillators \cite{lorenz} interacting directly through diffusive coupling and indirectly through a common environment $s$, which is modeled as a damped dynamical system \cite{resmiad, resmienv, tanenv, tanpre3}. The mathematical model of the coupled system is given by
\begin{subequations}
\label{evlsys}
\begin{align}
\label{lx}
\dot{x}_{1,2} &= \sigma(y_{1,2}-x_{1,2})+d(x_{2,1}-x_{1,2})+{\epsilon}s,\\
\label{ly}
\dot{y}_{1,2} &= (r-z_{1,2})x_{1,2}-{y}_{1,2},\\
\label{lz}
\dot{z}_{1,2} &= x_{1,2}y_{1,2}-bz_{1,2},\\
\label{ls}
\dot{s}&= -ks-\frac{{\epsilon}(x_1+x_2)}{2}.
\end{align}
\end{subequations}
Here $d$ is the diffusive coupling strength, and $\epsilon$ is the environmental coupling strength that controls the mutual interaction between the systems and environment. $k$ is the damping factor of the environment ($k>0$) \cite{resmiad}. This coupling scheme was proposed by \citet{resmiad} as a general scheme to introduce AD in any  coupled oscillators. Although Ref.~\cite{resmiad} studied the response of a Lorenz system under this coupling scheme, but it could not identify the OD state; only AD was reported there. Later the present authors \cite{tanpre3} reported that this coupling scheme can also gives rise to OD and AD-OD transitions in nonlinear oscillators. Thus, it is a generic coupling scheme to introduce AD and OD. Here, in the following, we investigate the effect of this coupling scheme in inducing the mixed mode oscillation suppression states in the Lorenz system.

Equation~(\ref{evlsys}) has a trivial homogeneous steady state (HSS), which is the origin $(0,0,0,0,0,0,0)$, and additionally two more coupling dependent nontrivial fixed points given by:

(i) $\mathcal{F}_{MMSS}\equiv(x^\dagger, y^\dagger, z^\dagger, -x^\dagger, -y^\dagger, z^\dagger, s^\dagger)$, where $x^\dagger= \pm \sqrt{\frac{\{(r-1)\sigma -2d\}b}{\sigma +2d}}$, $y^\dagger=\frac{(\sigma +2d)x^\dagger}{\sigma}$, $z^\dagger=\frac{(\sigma +2d){x^\dagger}^2}{\sigma b}$, $s^\dagger=0$. 
 $\mathcal{F}_{MMSS}$ gives the {\it mixed mode steady states} (MMSS) as for these steady state we have inhomogeneity in $x$ and $y$ variables (i.e.  $\pm x_1=\mp x_2, \pm y_1=\mp y_2$) and homogeneity in the $z$ variable  (i.e., $z_1=z_2$) . The {\it stabilization} of $\mathcal{F}_{MMSS}$ results in the {\it mixed mode death} (MMD) state, because here OD occurs in  $x$ and $y$ variables and AD occurs in the $z$ variable. Note that $\mathcal{F}_{MMSS}$ depends only upon $d$ and independent of $\epsilon$ and  $k$.

(ii) $\mathcal{F}_{NHSS}\equiv(x^\ast, y^\ast, z^\ast, x^\ast, y^\ast, z^\ast, s^\ast)$, where $x^\ast=\pm \sqrt{\frac{\{k\sigma(r-1)-{\epsilon}^2\}b}{k\sigma +{\epsilon}^2}}$, $y^\ast=\frac{(k\sigma +{\epsilon}^2)x^\ast}{k\sigma}$, $z^\ast=\frac{(k\sigma +{\epsilon}^2){x^\ast}^2}{k\sigma b}$, $s^\ast=-\frac{\epsilon x^\ast}{k}$. $\mathcal{F}_{NHSS}$ represents {\it nontrivial homogeneous steady states} (NHSS), stabilization of which gives rise to a novel nontrivial amplitude death (NAD) state (observed in $x$ and $y$ variables with $\pm x_1=\pm x_2, \pm y_1=\pm y_2$), which is a nonzero {\it bistable} state and a monostable AD state (observed in the $z$ variable with $z_1=z_2$)--we refer this state as mixed NAD and AD state, i.e., the MNAD state. It can be seen that, $\mathcal{F}_{NHSS}$ depends upon $\epsilon$ and $k$, and independent of $d$. 

As we notice here the symmetry of the system plays an important role behind the birth of $\mathcal{F}_{MMSS}$ and $\mathcal{F}_{NHSS}$, and thus the emergence of the MMD and MNAD states. The uncoupled Lorenz oscillators, denoted by, say, $\mathcal{L}_1$ and $\mathcal{L}_2$ are invariant under the reflection about the $z$-axis and Eqs.~ \eqref{evlsys} show the presence of a {\it translational symmetry} between the two Lorenz oscillators $\mathcal{L}_1$ and $\mathcal{L}_2$ under this coupling scheme. In the MNAD state, the relation between the oscillator variables clearly shows that the translational symmetry between $\mathcal{L}_1$ and $\mathcal{L}_2$ is preserved [$V_1=V_2$ ($V \equiv x ,~y$), $z_1=z_2$], however, the {\it reflection symmetry} about the $z$-axis is collapsed, as now $V_{1,2}$ and $-V_{1,2}$ are two different states. On the other hand, in the MMD state, the relation between the variables of $\mathcal{L}_1$ and $\mathcal{L}_2$ shows that $V_2=-V_1$, which indicates the destruction of translational symmetry, and also, since now  $V_{1,2}$ and $-V_{1,2}$ represent two different states, we can say that the reflection symmetry about the $z$-axis is also broken. Thus, in the MMD state both the translational symmetry and the reflection symmetry are broken.
\begin{figure}[t!]
\includegraphics[width=.45\textwidth]{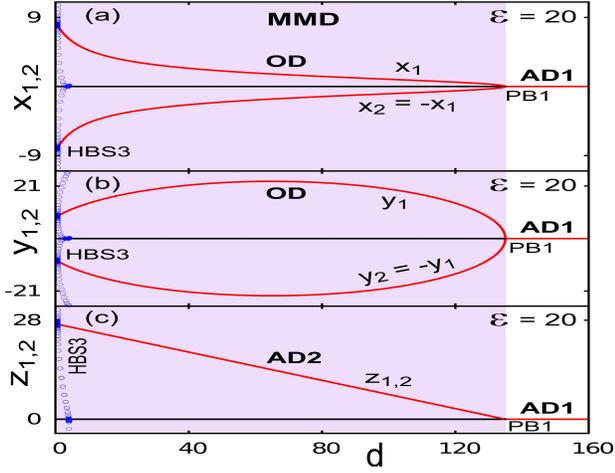}
\caption{\label{e-20} (Color online) Bifurcation diagram with variation of $d$ (using $\mbox{MATCONT}$) of two  identical chaotic Lorenz systems under direct-indirect coupling ($\epsilon = 20$, i.e., $\epsilon>\epsilon_{PB2}$). Grey (red) lines: Stable fixed points, Black lines: Unstable fixed points; open circle (blue): Unstable limit cycle. x and y variables show OD but the z variable shows AD (AD2). The MMD state (OD+AD2) (shown in shade) is created from AD1 through a supercritical pitchfork bifurcation at PB1 with decreasing d and destroyed at HBS3 (subcritical Hopf bifurcation point). Other parameters are $k=1$, $r=28$, $\sigma=10$, $b=\frac{8}{3}$.}
\end{figure}

Next, we theoretically analyze the stability of the system in order to explore the bifurcation scenarios. The characteristic equation of the system at the trivial HSS, $(0,0,0,0,0,0,0)$, is given by,
\begin{equation}
\label{trieigen}
(b+\lambda )^2\mathbf{F^2_T}({\lambda})\mathbf{F^3_T}({\lambda})=0,
\end{equation}
where $\mathbf{F^2_T}({\lambda})=({\lambda}^2+P'_{T1}\lambda +P'_{T0})$, $\mathbf{F^3_T}({\lambda})=({\lambda}^3+P_{T2}{\lambda}^2+P_{T1}\lambda +P_{T0})$ with $P'_{T1}=(1+\sigma +2d)$, $P'_{T0}=\sigma (1-r)+2d$, $P_{T2}=(1+\sigma +k)$, $P_{T1}=\sigma(1-r)+k(1+\sigma)+{\epsilon}^2$, and $P_{T0}={\epsilon}^2+K\sigma(1 - r)$. 
Since Eq.\eqref{trieigen} is a {\it seventh-order} polynomial, it is difficult to derive the exact analytical expressions of all the eigenvalues. But we can  predict the stability scenario of the trivial HSS from the properties of the coefficients of the characteristic equation itself \cite{math}. We find that all the eigenvalues at the trivial HSS have negative real part and thus give rise to AD1, when $k>-(1+\sigma)$, $d>\frac{\sigma(r-1)}{2}$ and $\epsilon >\sqrt{k\sigma(r-1)}$. With {\it decreasing} strength of the coupling parameters ($d$ and $\epsilon$) we observe different dynamical behaviors in the coupled system. A close inspection of the nontrivial fixed points reveals that $\mathcal{F}_{MMSS}$ and $\mathcal{F}_{NHSS}$ appear through a pitchfork bifurcation at $d_{PB1}$ and ${\epsilon}_{PB2}$, respectively, where
\begin{equation}
\label{dpb}
d_{PB1}=\frac{(r-1)\sigma}{2}, 
\end{equation}
\begin{equation}
\label{epb}
{\epsilon}_{PB2}=\sqrt{k\sigma(r-1)}.
\end{equation}
To keep the uncoupled Lorenz systems in the chaotic region, we set the system parameters at $r=28$, $b=\frac{8}{3}$ and $\sigma = 10$ throughout this paper. 

Figure~\ref{e-20} shows the one-dimensional bifurcation diagram with variable $d$ using the $\mbox{MATCONT}$ package \cite{mat} with an exemplary value  $k=1$ and $\epsilon = 20$ (note that $\epsilon>\epsilon_{PB2}=16.432$). It can be clearly seen that $x_{1,2}$ and $y_{1,2}$ variables show the OD state [Figs.~\ref{e-20} (a) and \ref{e-20} (b)], whereas, the $z_{1,2}$ variables show an AD state (AD2) [Fig.~\ref{e-20}(c)]; both these OD and AD states emerge from the amplitude death state (AD1) through a supercritical pitchfork bifurcation at $d_{PB1}=135$ [this value exactly matches with Eq.\eqref{dpb}].  Thus, for $\epsilon=20$, a decreasing $d$ gives rise to an MMD state (OD+AD2) from an AD state (AD1). The MMD state becomes unstable through a subcritical Hopf bifurcation (HBS3). 
\begin{figure}[t!]
\includegraphics[width=.45\textwidth]{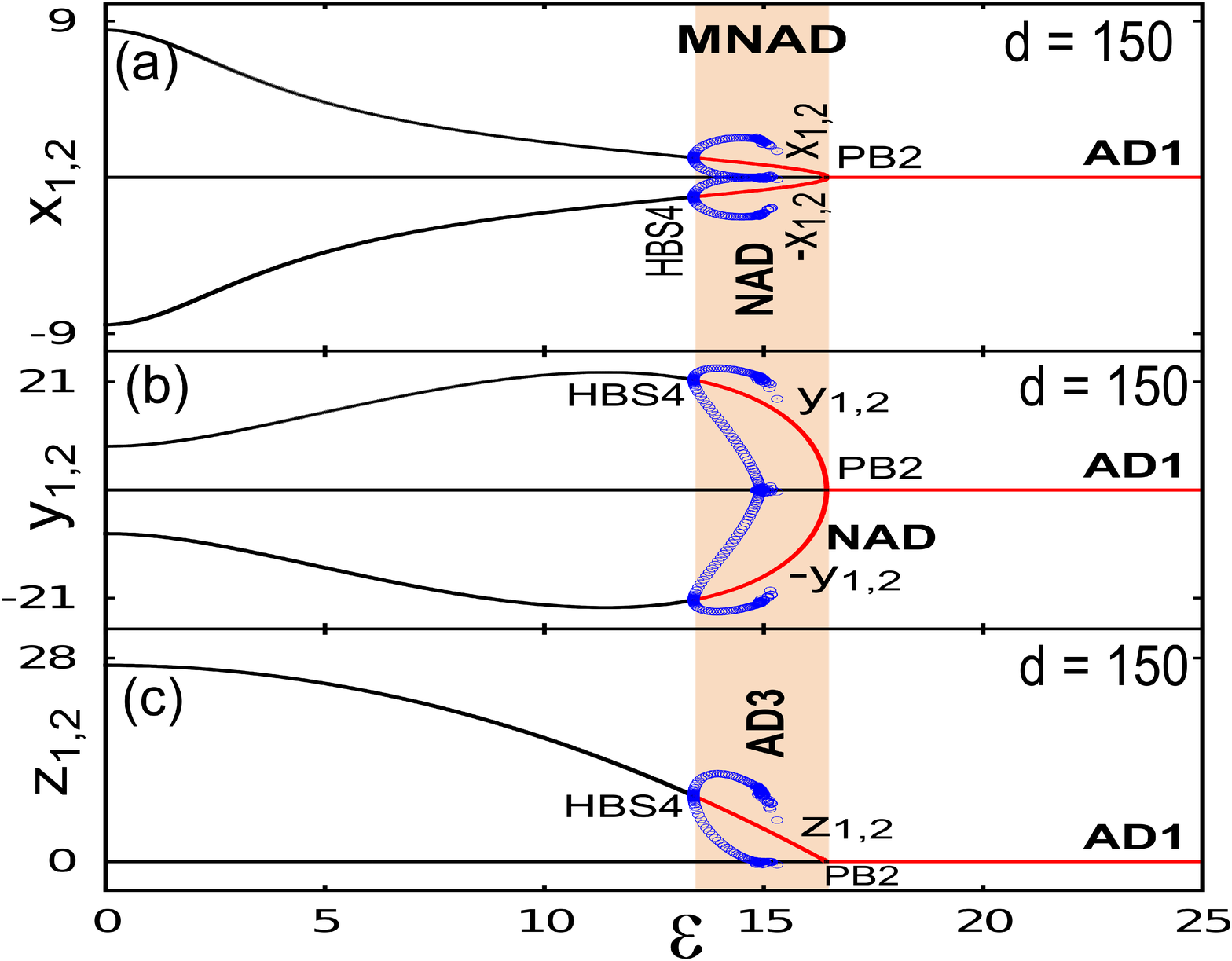}
\caption{\label{d-150} (Color online) Bifurcation diagram of two  identical chaotic Lorenz systems under direct-indirect coupling with variation of $\epsilon$. x and y show the NAD state and z shows an AD state (AD3). The MNAD state (shown in shade) emerges from AD1 at PB2 (supercritical pitchfork bifurcation point) with decreasing $\epsilon$ ($d=150$, i.e., $d>d_{PB1}$) and destroyed at HBS4 (subcritical Hopf bifurcation point). Other parameters are same as Fig.~\ref{e-20}.}
\end{figure}

Next, we fix the value of $d$ at $d=150$ ($d>d_{PB1}$), and vary $\epsilon$. Figure~\ref{d-150} shows that, with {\it decreasing} $\epsilon$, the bistable NAD state occurs in $x_{1,2}$ and $y_{1,2}$ [Figs.~\ref{d-150}(a) and \ref{d-150}(b)] and an AD state (AD3) occurs in $z_{1,2}$ [Fig.~\ref{d-150}(c)] at $\epsilon_{PB2}=16.432$ [this value exactly matches with Eq.\eqref{epb}]; both these states emerge from the AD1 state. Thus, a decreasing $\epsilon$ (and a proper value of $d$) gives rise to an MNAD state (NAD+AD3) from the AD1 state. The MNAD state is destroyed through a subcritical Hopf-bifurcation (HBS4).
\begin{figure}
\includegraphics[width=.45\textwidth]{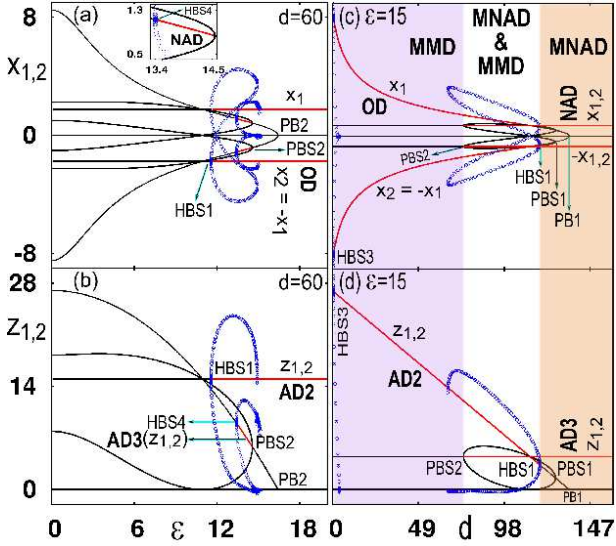}
\caption{\label{d60e15} (Color online) Between HBS4 and PBS2 (a) OD and NAD and (b) AD2 and AD3 coexist. (a) Inset shows the zoomed in view of the NAD state. (c) NAD and OD and (d) AD3 and AD2 coexist between PBS2 and HBS1. HBS and PBS represent the subcritical Hopf and subcritical pitchfork bifurcation points, respectively. Other parameters are same as in Fig.~\ref{e-20}.}
\end{figure}

To identify several other bifurcation curves that mark the distinct regions of occurrence of oscillation suppression states and their coexistence we consider the characteristic equation corresponding to the nontrivial fixed points ($x^i, y^i, z^i, Jx^i, Jy^i, z^i, s^i$), where $J = \pm 1$ and $i = \ast~\mbox{or}~\dagger$, which is given by
\begin{equation}
\label{nteigen}
\mathbf{F^3_N}({\lambda})\mathbf{F^4_N}({\lambda})=0,
\end{equation}
where $\mathbf{F^3_N}({\lambda})=({\lambda}^3+P'^i_{N2}{\lambda}^2+P'^i_{N1}{\lambda}+P'^i_{N0})$, and $\mathbf{F^4_N}({\lambda})=({\lambda}^4+P^i_{N3}{\lambda}^3+P^i_{N2}{\lambda}^2+P^i_{N1}\lambda +P^i_{N0})$ with $P'^i_{N2}=\sigma +2d+b+1$, $P'^i_{N1}= b+(b+1)(\sigma +2d)+{x^i}^2+\sigma(z^i-r)$, $P'^i_{N0}=(\sigma +2d)(b+{x^i}^2)+\sigma \{x^iy^i+b(z^i-r)\}$, $P^i_{N3}=k+b+1+\sigma$, $P^i_{N2}=k(b+1+\sigma)+\sigma(b+1-r+z^i)+b+{x^i}^2+{\epsilon}^2$, $P^i_{N1}=k\{\sigma(b+1-r+z^i)+b+{x^i}^2\}+\sigma\{b(1-r+z^i)+x^i(x^i+y^i)\}+{\epsilon}^2(b+1)$ and $P^i_{N0}={\epsilon}^2(b+{x^i}^2)+k\sigma\{b(1-r+z^i)+x^i(x^i+y^i)\}$. From Eq.~(\ref{nteigen}) with $J=1$ and $i=\ast$ (i.e., for the fixed point $\mathcal{F}_{NHSS}$) we find that $d$ appears only in the term $\mathbf{F^3_N}({\lambda})$, i.e., it controls only three eigenvalues. Similarly from Eq.~(\ref{nteigen}) with $J=-1$ and $i=\dagger$ (i.e., for the fixed point $\mathcal{F}_{MMSS}$) we conclude that behavior of the four eigenvalues are controlled by $\epsilon$ and $k$ as they appear only in the term $\mathbf{F^4_N}({\lambda})$.

For $d<d_{PB1}$, $k=1$ and $\epsilon_{PBS2}>\epsilon_{HBS4}$, the MNAD state appears through a pitchfork bifurcation at PBS2. The analytical expression is obtained by putting $P'^i_{N0}=0$ \cite{math} and is given by
\begin{equation}
\label{pbs2}
d_{PBS2} = \frac{-\sigma\{ b(1+z^\ast -r)+x^\ast(x^\ast +y^\ast)\}}{2(b+{x^\ast}^2)}.
\end{equation}
Another Hopf bifurcation point HBS2 is observed for lower values of $d$. To derive the locus we set $(P'^i_{N1}P'^i_{N2}-P'^i_{N0})=0$ \cite{math} and get 
\begin{equation}
\label{hbs2}
d_{HBS2} = \frac{-B^\ast+\sqrt{{B^\ast}^2-4{A^\ast}C^\ast}}{2A^\ast},
\end{equation} 
where, $A^\ast = 4(b+1)$, $B^\ast = 2(b+1)(2\sigma +b+1)+2\sigma(z^\ast -r)$, $C^\ast = {\sigma}^2(b+1-r+z^\ast)+\sigma \{ (b+1)^2-r+z^\ast- x^\ast y^\ast \}+(b+{x^\ast}^2)(b+1)$. When $d<d_{PB}$, depending upon the choice of $k$ and $d$ values, the MNAD state appears either through PBS2 or HBS2. Figures~\ref{d60e15}(a, b) show that for $d=60$, i.e., $d<d_{PB1}$ and $k=1$, the AD1 state disappears; coexistence of NAD (AD3) and OD (AD2) is observed between PBS2 and HBS4. For ${\epsilon}<{\epsilon}_{PB2}$, the AD1 state vanishes and we get two possible routes to the MMD state: (i) Pitchfork bifurcation (PBS1) and (ii) Hopf bifurcation (HBS1). With the proper choice of $k$ and $\epsilon$ we can select one of these routes to the MMD state. Figures~\ref{d60e15}(c, d) show the Hopf bifurcation route to MMD for $\epsilon = 15$, i.e., ${\epsilon}<{\epsilon}_{PB2}$ and $k=1$. To get the exact locus of PBS1 and HBS1 we set $P^i_{N0}=0$ and $(P^i_{N1}P^i_{N2}P^i_{N3}-{P^i_{N1}}^2-P^i_{N0}{P^i_{N3}}^2)=0$, respectively, and get
\begin{subequations}
\label{pbs1hbs1}
\begin{align}
\label{pbs1}
\epsilon_{PBS1}&=\sqrt{\frac{k\sigma \{b(r-1-z^\dagger)-x^\dagger(x^\dagger +y^\dagger)\}}{b+{x^\dagger}^2}},\\
\label{hbs1}
\epsilon_{HBS1}&=\sqrt{\frac{-B^\dagger +\sqrt{{B^\dagger}^2-4A^\dagger C^\dagger}}{2A^\dagger}},
\end{align}
\end{subequations}
where, $A^\dagger = (b+1)(k+\sigma)$, $B^\dagger = (b+1)[{P^\dagger}_{N3}({P^\dagger}_{N2}-{\epsilon}^2)-2\{{P^\dagger}_{N1}-{\epsilon}^2(b+1)\}]+{P^\dagger}_{N3}[\{{P^\dagger}_{N1}-{\epsilon}^2(b+1)\}-{P^\dagger}_{N3}(b+{x^\dagger}^2)]$, $C^\dagger = {P^\dagger}_{N3}[({P^\dagger}_{N2}-{\epsilon}^2)\{{P^\dagger}_{N1}-{\epsilon}^2(b+1)\}-{P^\dagger}_{N3}\{{P^\dagger}_{N0}- {\epsilon}^2(b+{x^\dagger}^2)\}]-\{{P^\dagger}_{N1}-{\epsilon}^2(b+1)\}^2$. 
\begin{figure}
\includegraphics[width=.4\textwidth]{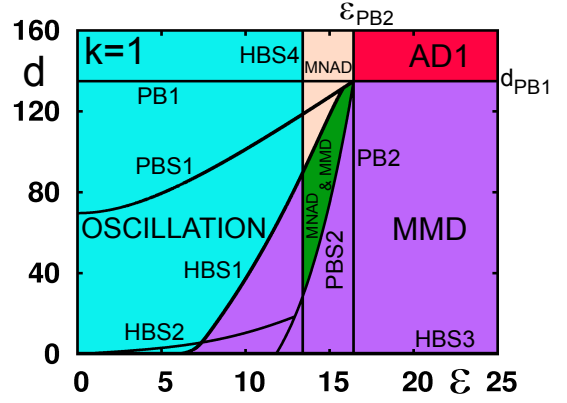}
\caption{\label{detpbd} (Color online) Two-parameter bifurcation diagram (using $\mbox{XPPAUT}$) in $\epsilon-d$ space. Other parameters: $k=1$, $r=28$, $\sigma = 10$ and $b=\frac{8}{3}$.}
\end{figure}

The loci of HBS4 and HBS3 could not be found in the closed form; thus, to present a complete bifurcation scenario we compute the two parameter bifurcation diagram (Fig.~\ref{detpbd}) in the $\epsilon-d$ space with $k=1$ [i.e., with $k>-(1+\sigma)$] using the $\mbox{XPPAUT}$ package \cite{xpp}, which  exactly agree with our theoretically obtained bifurcation curves. 

\subsection{Identical Lorenz systems interacting through mean-field diffusive coupling}
\label{sec2b}
\noindent To verify that the mixed mode oscillation suppression states are not limited to the direct-indirect coupling only, we consider another generic coupling scheme in the context of AD and OD, namely the mean-field diffusive coupling \cite{tanpre1,*tanpre2}, and investigate the occurrence of MMD and MNAD states. We consider the following two identical mean-field coupled Lorenz systems:
\begin{subequations}
\label{mflsys}
\begin{align}
\label{mlx}
\dot{x}_{1,2} &= \sigma(y_{1,2}-x_{1,2})+\epsilon \left [\frac{Q(x_1+x_2)}{2} -x_{1,2}\right],\\
\label{mly}
\dot{y}_{1,2} &= (r-z_{1,2})x_{1,2}-{y}_{1,2},\\
\label{mlz}
\dot{z}_{1,2} &= x_{1,2}y_{1,2}-bz_{1,2}.
\end{align}
\end{subequations}
Here $\epsilon$ is the coupling strength and the control parameter $Q$ determines the density of mean-field ($0\leq Q<1$) \cite{qstr}. From Eqs.~(\ref{mflsys}) we can see that the origin ($0,0,0,0,0,0$) is the homogeneous steady state (HSS). Also, we have two more coupling-dependent nontrivial fixed points: 
(i) $\mathcal{F}_{MMSS}\equiv(x^\dagger, y^\dagger, z^\dagger, -x^\dagger, -y^\dagger, z^\dagger)$, where $x^\dagger = \pm \sqrt{\frac{b(\sigma r-\sigma -\epsilon)}{\sigma +\epsilon}}$, $y^\dagger = \frac{x^\dagger(\sigma+\epsilon)}{\sigma}$, $z^\dagger = \frac{{x^\dagger}^2(\sigma+\epsilon)}{b\sigma}$. 
(ii) $\mathcal{F}_{NHSS}\equiv(x^\ast, y^\ast, z^\ast, x^\ast, y^\ast, z^\ast)$, where $x^\ast = \pm \sqrt{\frac{b\{\sigma(r-1)-\epsilon(1-Q)\}}{\sigma + \epsilon(1-Q)}}$, $y^\ast = \frac{x^\ast\{\sigma+\epsilon(1-Q)\}}{\sigma}$, $z^\ast = \frac{{x^\ast}^2\{\sigma + \epsilon(1-Q)\}}{b\sigma}$.

The nontrivial fixed points $\mathcal{F}_{MMSS}$ and $\mathcal{F}_{NHSS}$ emerge due to the symmetry breaking pitchfork bifurcations at $\epsilon_{PB2}$ and $\epsilon_{PB1}$, respectively,
\begin{subequations}
\label{mflpb}
\begin{align}
\label{mflpb1}
\epsilon_{PB1} &= \frac{\sigma(r-1)}{1-Q},\\
\label{mflpb2}
\epsilon_{PB2} &= \sigma(r-1).
\end{align}
\end{subequations} 
\begin{figure}[t!]
\includegraphics[width=.4\textwidth]{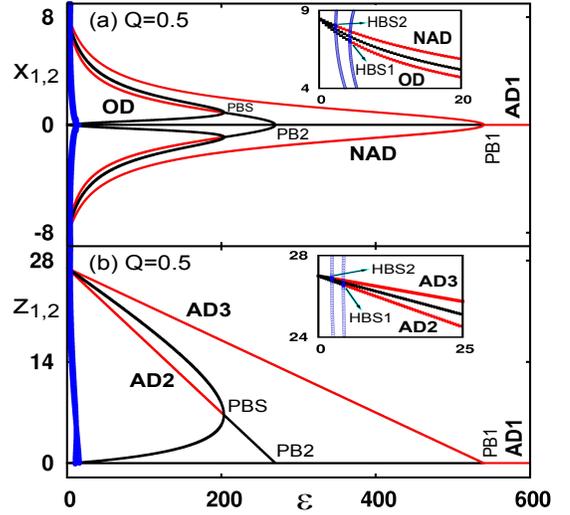}
\caption{\label{mf_lorenz} (Color online) Bifurcation diagram of two identical chaotic Lorenz attractors coupled through mean-field diffusion at $Q=0.5$. Here MNAD (NAD+AD3) is created at PB1; The MMD state (OD+AD2) created at PBS is always accompanied by the MNAD state; PBS: Subcritical pitchfork bifurcation. (a) In $x$ variable: AD1 to NAD transition (at PB1) and co-existence of OD and NAD  (between PBS and HBS1). Same is for the $y$ variable (not shown here). (b) In $z$ variable: AD1 to AD3 transition (at PB1) and co-existence of AD2 and AD3 (between PBS and HBS1) is observed. ($r=28$, $\sigma=10$, $b=\frac{8}{3}$).}
\end{figure}

To explore the complete bifurcation scenario, we write the characteristic equation of the system at the nontrivial fixed points ($x^i, y^i, z^i, Jx^i, Jy^i, z^i$), where $J = \pm 1$ and $i = \ast~\mbox{or}~\dagger$, as
\begin{equation}
\label{mflce}
\mathbf{F}'^3_N({\lambda})\mathbf{F}''^3_N({\lambda})=0,
\end{equation}
where, $\mathbf{F}^{l3}_N({\lambda})= ({\lambda}^3+U^{li}_{N2}{\lambda}^2+U^{li}_{N1}\lambda +U^{li}_{N0})$ with $l=~'\mbox{or}~''$, $U'^i_{N2}=b+1+\sigma +\epsilon(1-Q)$, $U'^i_{N1}=b+(b+1)(\sigma +\epsilon - \epsilon Q)+{x^i}^2-\sigma(r-z^i)$, $U'^i_{N0}=\sigma x^iy^i-b\sigma(r-z^i)+(\sigma +\epsilon -\epsilon Q)({x^i}^2+b)$, $U''^i_{N2}=\sigma +\epsilon +b+1$, $U''^i_{N1}=(\sigma +\epsilon)(b+1)+b+{x^i}^2+\sigma(z^i-r)$, $U''^i_{N0}=(\sigma +\epsilon)(b+{x^i}^2)+\sigma x^iy^i+b\sigma(z^i-r)$.

Using a similar approach adopted in the Sec.~\ref{sec2a}, we derive the locus of the bifurcation curves and they are given by
\begin{subequations}
\label{mflnt}
\begin{align}
\label{mflpbs}
\epsilon_{PBS} &= \frac{\sigma[rQ^{\prime}-4+\sqrt{r^2Q^{\prime 2}+8Qr}]}{4},\\
\label{mflhbs1}
Q_{HBS1} &= \frac{-B^{\dagger}_1-\sqrt{{B^{\dagger}_1}^2-4A^{\dagger}_1C^{\dagger}_1}}{2A^{\dagger}_1}.
\end{align}
\end{subequations}
Here $Q^{\prime}=(2-Q)$, $A^{\dagger}_1={\epsilon}^2(b+1)$, $B^{\dagger}_1=-(\sigma +\epsilon)[ \frac{\epsilon {x^\dagger}^2}{b}+2\epsilon(b+1)]+\sigma r \epsilon - (b+1)^2\epsilon$, $C^{\dagger}_1=[ b+1-\sigma-\epsilon+\frac{(1+\sigma+\epsilon)(\sigma+\epsilon)}{b}]{x^\dagger}^2-\sigma r(1+\sigma+\epsilon)+b(b+1)+(b+1)^2(\sigma+\epsilon)+(b+1)(\sigma+\epsilon)^2$. Here $\epsilon_{PBS}$ gives the coupling strength at which the MMD state emerges (due to the stabilization of $\mathcal{F}_{MMSS}$) through a subcritical pitchfork bifurcation. From Eq.~\eqref{mflpb2} and Eq.~\eqref{mflpbs} it is clear that although the emerging point of $\mathcal{F}_{MMSS}$ is independent of $Q$, but its stabilization, i.e., the creation of the MMD state, is controlled by $Q$. 

Figures~\ref{mf_lorenz}(a) and \ref{mf_lorenz}(b) show the bifurcation diagram of $x_{1,2}$ and $z_{1,2}$, respectively for $Q=0.5$ [$y_{1,2}$ behaves in a similar way as $x_{1,2}$ and thus not shown in the figure]. With this coupling scheme we obtain all the oscillation quenching states, namely the MMD (i.e., OD + AD2) and the MNAD (i.e., NAD + AD3) state. It is noteworthy that the MMD state is always accompanied by the MNAD state. Also, here the AD1 to MMD transition is not possible, as for any $Q>0$ one has ${\epsilon}_{PB1}>{\epsilon}_{PB2}$. However, the direct transition from AD1 to MNAD takes place at ${\epsilon}_{PB1}$.

\begin{figure}
\includegraphics[width=.35\textwidth]{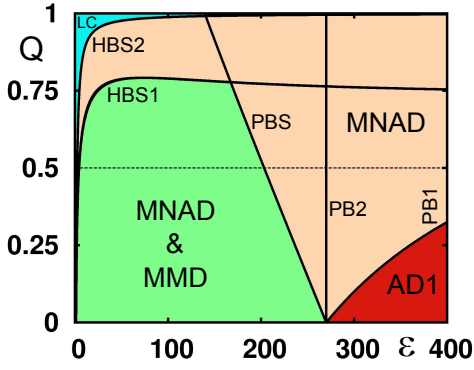}
\caption{\label{tpbd_mflorenz} (Color online) Two-parameter bifurcation diagram of two identical chaotic Lorenz systems coupled through mean-field diffusion. The horizontal dotted line indicates the $Q$ value for which Fig.~\ref{mf_lorenz} is drawn ($r=28$, $\sigma=10$, $b=\frac{8}{3}$).}
\end{figure}

The complete bifurcation scenario is shown in Fig.~\ref{tpbd_mflorenz} in the $\epsilon-Q$ space. Here the intersection of HBS1 curve with PBS organizes the coexisting MMD and MNAD state, and the region bounded by HBS2 and PB1 curve organizes the occurrence of the MNAD state. The horizontal dotted line indicates the density of the mean field for which Fig.~\ref{mf_lorenz} is drawn.
\section{CHEN system}
\label{sec3}
\subsection{Identical Chen systems interacting through direct-indirect coupling}
\label{sce3a}
\noindent Next, we verify the generality of the occurrence of the mixed mode oscillation suppression states in a chaotic Lorenz-like system, namely the Chen system \cite{chen}. Mathematical model of two identical Chen systems under the direct-indirect coupling scheme is given by
\begin{subequations}
\label{envchen}
\begin{align}
\label{cx}
\dot{x}_{1,2} &= a({y}_{1,2}-{x}_{1,2})+d(x_{2,1}-{x}_{1,2})+{\epsilon}s,\\
\label{cy}
\dot{y}_{1,2} &= (c-a){x}_{1,2}-{x}_{1,2}{z}_{1,2}+c{y}_{1,2},\\
\label{cz}
\dot{z}_{1,2} &= x_{1,2}y_{1,2}-bz_{1,2},\\
\label{cs}
\dot{s} &= -ks-\frac{{\epsilon}(x_1+x_2)}{2}.
\end{align}
\end{subequations}
Here $a>0$, $c$ ($2c>a$) and $b>0$ are the system parameters. In addition to the trivial homogeneous steady state (HSS), i.e., the origin (0,0,0,0,0,0,0), the system has  two more coupling dependent nontrivial fixed points (i) $\mathcal{F}_{MMSS}\equiv(x^\dagger, y^\dagger, z^\dagger, -x^\dagger, -y^\dagger, z^\dagger, s^\dagger)$, where $x^\dagger= \pm \sqrt{\frac{b\{2c(a+d)-a^2\}}{a+2d}}$, $y^\dagger=\frac{x^\dagger(a+2d)}{a}$, $z^\dagger=\frac{{x^\dagger}^2(a+2d)}{ab}$, $s^\dagger=0$ and (ii) $\mathcal{F}_{NHSS}\equiv(x^\ast, y^\ast, z^\ast, x^\ast, y^\ast, z^\ast, s^\ast)$, where $x^\ast= \pm \sqrt{\frac{b\{ka(c-a)+c(ka+{\epsilon}^2)\}}{ka+{\epsilon}^2}}$, $y^\ast=\frac{x^\ast(ka+{\epsilon}^2)}{ka}$, $z^\ast=\frac{{x^\ast}^2(ka+{\epsilon}^2)}{kab}$, $s^\ast=-\frac{\epsilon x^\ast}{k}$. A close inspection of the nontrivial fixed points reveals that
\begin{equation}
d_{PB} = \frac{a^2-2ac}{2c},
\end{equation}
where $d_{PB}$ gives the coupling strength at which $\mathcal{F}_{MMSS}$ emerges.

\begin{figure}
\includegraphics[width=.45\textwidth]{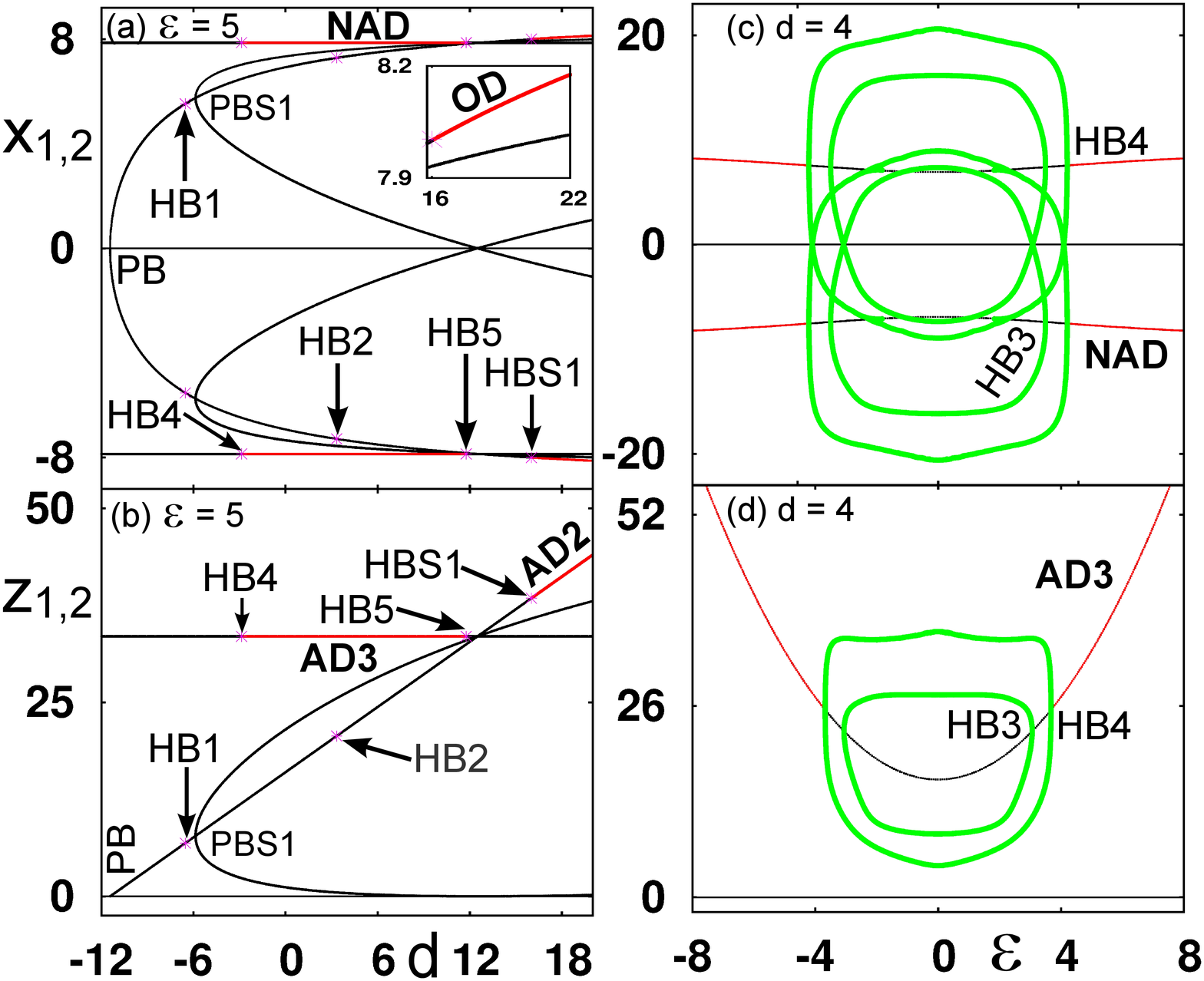}
\caption{\label{chen-e-5} (Color online) Bifurcation diagram of two identical chaotic Chen systems coupled through direct-indirect coupling. (a,b) With $\epsilon = 5$, the MMD state (OD+AD2) born through subcritical Hopf bifurcation (HBS1). NAD (AD3) state exists between HB5 and HB4. (a) Inset shows the zoomed in view of the OD state. (c,d) With $d = 4$ MNAD is noticed. (c) NAD and (d) AD3 state appear at HB4. PB, PBS, HB and HBS denote the pitchfork, subcritical pitchfork, Hopf, subcritical Hopf bifurcation points, respectively. Other parameters are $k=1$, $a=40$, $c=28$, $b=3$.}
\end{figure}

From the detailed analysis of the characteristic equation we get the stability condition of the trivial HSS as $k>(c-a)$, ${\epsilon}^2<\frac{k(a^2-2ac)}{c}$ and $d<\frac{a^2-2ac}{2c}$. Considering the conditions for the system parameters (i.e.  $a>0$, $2c>a$ and $b>0$) of the chaotic Chen attractor we get $(a^2-2ac)<0$ for all possible set of parameter values. So for any positive $k$,  the stability condition for the trivial HSS is satisfied when ${\epsilon}^2<0$, i.e., for imaginary $\epsilon$ values. These conditions clearly show that for any positive $k$, the trivial HSS remains always unstable and the AD state, that arises due to stabilization of the trivial HSS, never appears. The detailed bifurcation scenario of the system is shown in Fig.~\ref{chen-e-5} with an exemplary value $k=1$.  Figures~\ref{chen-e-5}(a) and \ref{chen-e-5}(b) show the bifurcation structure for $\epsilon = 5$; here we can see the presence of MMD (OD+AD2) and MNAD (NAD+AD3). To show the complete bifurcation structure we also consider the negative $d$ values (and later, also negative $\epsilon$ values). Figures~\ref{chen-e-5} (c) and \ref{chen-e-5} (d) show the bifurcation for varying $\epsilon$ and fixed $d = 4$; it shows the presence of MNAD (NAD+AD3) state in the coupled system. 
\subsection{Identical Chen systems interacting through mean-field diffusive coupling}
\label{sec3b}

\begin{figure}[t!]
\includegraphics[width=.35\textwidth]{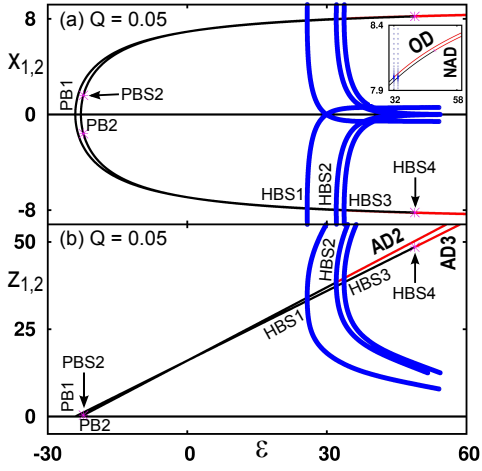}
\caption{\label{mf_chen} (Color online) Bifurcation diagram of two identical chaotic Chen systems coupled through mean-field diffusion. The stabilization of  $\mathcal{F}_{MMSS}$ ($\mathcal{F}_{NHSS}$) gives rise to (a) OD (NAD)  and (b) AD2 (AD3) state. Other parameters are $Q=0.05$, $a=40$, $c=28$, $b=3$.}
\end{figure}
\noindent Next, we verify the occurrence of MMD and MNAD states in two identical Chen systems under the mean-field diffusive coupling scheme. The mathematical model of the coupled system is given by
\begin{subequations}
\label{mfchen}
\begin{align}
\label{mfcx}
\dot{x}_{1,2} &= a({y}_{1,2}-{x}_{1,2})+\epsilon \bigg(\frac{Q(x_1+x_2)}{2}-x_1\bigg),\\
\label{mfcy}
\dot{y}_{1,2} &= (c-a){x}_{1,2}-{x}_{1,2}{z}_{1,2}+c{y}_{1,2},\\
\label{mfcz}
\dot{z}_{1,2} &= x_{1,2}y_{1,2}-bz_{1,2}.
\end{align}
\end{subequations}
The Eq.~\eqref{mfchen} has the trivial fixed point ($0,0,0,0,0,0$) and two more coupling dependent nontrivial fixed points (i) $\mathcal{F}_{MMSS}\equiv(x^\dagger, y^\dagger, z^\dagger, -x^\dagger, -y^\dagger, z^\dagger)$ where $x^\dagger = \pm \sqrt{\frac{ab(c-a)+cb(a+\epsilon)}{a+\epsilon}}$, $y^\dagger = \frac{x^\dagger(a+\epsilon)}{a}$, $z^\dagger = \frac{{x^\dagger}^2(a+\epsilon)}{ab}$ and (ii) $\mathcal{F}_{NHSS}\equiv(x^\ast, y^\ast, z^\ast, x^\ast, y^\ast, z^\ast)$, where $x^\ast = \pm \sqrt{\frac{b\{2ac-a^2+c\epsilon(1-Q)\}}{a+\epsilon(1-Q)}}$, $y^\ast = \frac{x^\ast\{a+\epsilon(1-Q)\}}{a}$, $z^\ast = \frac{{x^\ast}^2\{a+\epsilon(1-Q)\}}{ab}$. $\mathcal{F}_{MMSS}$ and $\mathcal{F}_{NHSS}$ born through the pitchfork bifurcation at $\epsilon_{PB2}$ and $\epsilon_{PB1}$, respectively, where
\begin{equation}
\epsilon_{PB1} = \frac{a^2-2ac}{c(1-Q)},
\end{equation}
\begin{equation}
\epsilon_{PB2} = \frac{a^2-2ac}{c}.
\end{equation}
The stabilization of  $\mathcal{F}_{MMSS}$ and $\mathcal{F}_{NHSS}$ gives rise to the MMD and MNAD states, respectively. Figure~\ref{mf_chen}(a) and \ref{mf_chen}(b) show the bifurcation diagram of $x_{1,2}$ and $z_{1,2}$, respectively for $Q=0.05$. Both the MMD (OD+AD2) and MNAD (NAD+AD3) states appear through subcritical hopf bifurcation at HBS2 and HBS4, respectively.

\section{EXPERIMENT}
\label{sec4}
\begin{figure}[t!]
\includegraphics[width=.45\textwidth]{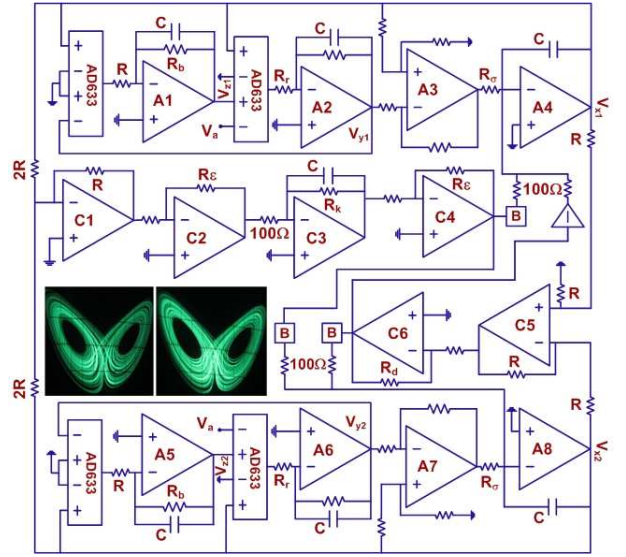}
\caption{\label{envckt} (Color online) Experimental circuit diagram of two coupled Lorenz systems under direct-indirect coupling. A1-A8, C1-C6 are realized with TL082 op-amps. All the unlabeled resistors have value $R_i=100$~k$\Omega$, C=10 nF, $V_a=3$ volt, $R=10$~k$\Omega$, $R_b=39$~k$\Omega$, $R_r=1.07$~k$\Omega$, $R_{\sigma}=10$~k$\Omega$, $R_k=100$~k$\Omega$. $\pm15$ volt power supplies are used; resistors (capacitors) have $\pm5\%$ ($\pm1\%$) tolerance. Box denoted by ``B" are op-amp based buffers; inverters are realized with the unity-gain non-inverting op-amps. Insets (in the middle part) shows the experimental attractors from the uncoupled Lorenz oscillators $\mathcal{L}_1$ (left) and $\mathcal{L}_2$ (right) ($y$-axis: 1 v/div, $x$-axis: 1 v/div).}
\end{figure}
\noindent We experimentally verify the occurrence of MMD and MNAD states in the identical Lorenz attractor interacting through direct-indirect coupling. To implement the practical electronic circuit we have rescaled \cite{lrnzckt} the variables of Eq.~\eqref{evlsys} using $x_i = (\sqrt{\frac{3}{r}})x_i$, $y_i = (\sqrt{\frac{3}{r}})y_i$, $z_i = (\frac{3}{r})z_i$, $s_i = (\sqrt{\frac{3}{r}})s_i$ where $i = 1,2$. Then the modified equations become
\begin{subequations}
\label{evlexpt}
\begin{align}
\label{lxe}
\dot{x}_{1,2} &= \sigma(y_{1,2}-x_{1,2})+d(x_{2,1}-x_{1,2})+{\epsilon}s,\\
\label{lye}
\dot{y}_{1,2} &= \frac{r}{3}(3-z_{1,2})x_{1,2}-{y}_{1,2},\\
\label{lze}
\dot{z}_{1,2} &= x_{1,2}y_{1,2}-bz_{1,2},\\
\label{lse}
\dot{s} &= -ks-\frac{{\epsilon}(x_1+x_2)}{2}.
\end{align}
\end{subequations}
Figure.~\ref{envckt} represents the electronic circuit of the coupled Lorenz systems \cite{lrnzckt} with the direct-indirect coupling given by Eqs.~\eqref{evlexpt}. A1-A4  and A5-A8 are used to realize the individual Lorenz oscillators $\mathcal{L}_1$ and $\mathcal{L}_2$, respectively. The subunit realized with op-amps C5-C6 produces the diffusive coupling part, and the subunit consists of op-amps C1-C4 mimics the environmental coupling part. The voltage equation of the circuit of Fig. \ref{envckt} can be written as follows:
\begin{figure}
\includegraphics[width=.45\textwidth]{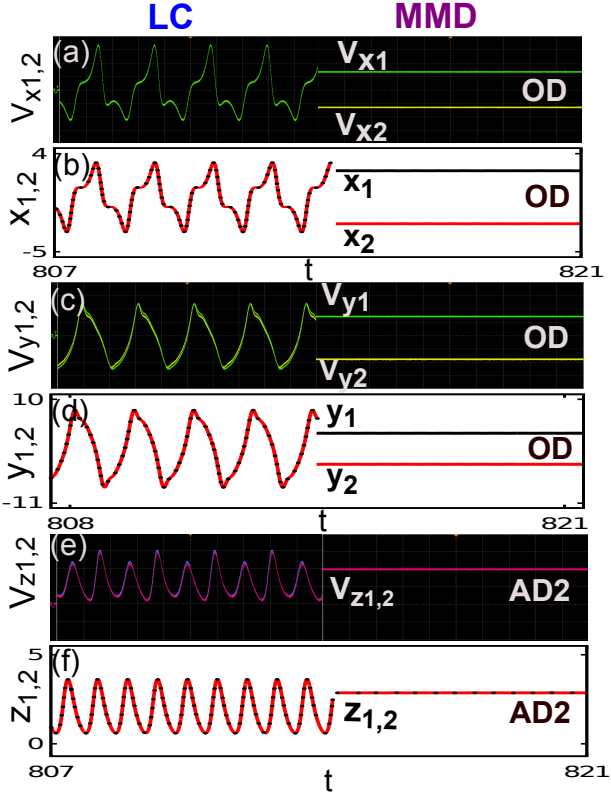}
\caption{\label{exptmmd} (Color online) {\bf Transition from limit cycle (LC) to MMD}: (a, c, e) Experimental real time traces of $V_{xi}$, $V_{yi}$ and $V_{zi}$ along with the (b, d, f) numerical time series plots  of $x_i$, $y_i$ and $z_i$ where $i=1,2$. $R_{\epsilon} = 1.127$~k$\Omega$ ($\epsilon = 11.27$): Left panels show LC for $R_d = 4.55$~k$\Omega$ ($d=45.5$). Right panels show MMD for $R_d = 145.6$~$\Omega$ ($d=1.456$). Both the experimental and numerical results clearly show that the MMD state is a mixed state of OD (observed in $x$, $y$ and $V_{x}$, $V_{y}$) and AD (AD2) state (observed in $z$ and $V_{z}$). For other parameters see text. Left panel, $y$ axis: (a) 100 mv/div (c) 340 mv/div (e) 128 mv/div; $x$ axis: 800 $\mu$s/div. Right panel, $y$ axis: (a) 100 mv/div (c) 300 mv/div (e) 128 mv/div; $x$ axis: 10 ms/div.}
\end{figure}
\begin{figure}
\includegraphics[width=.45\textwidth]{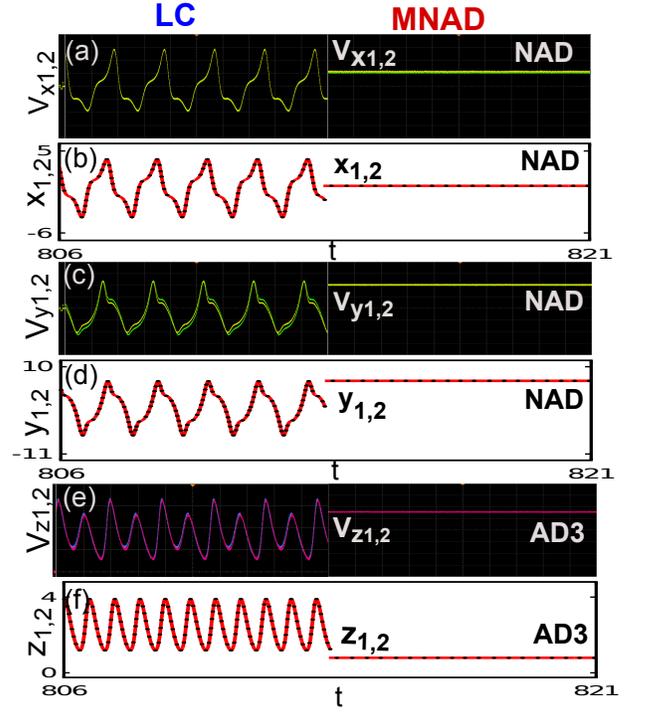}
\caption{\label{exptmnad} (Color online) {\bf Transition from limit cycle (LC) to MNAD}: (a, c, e) Experimental real time traces of $V_{xi}$, $V_{yi}$ and $V_{zi}$ along with the (b, d, f) numerical time series plots  of $x_i$, $y_i$ and $z_i$ where $i=1,2$. $R_d = 8.5$~k$\Omega$ ($d=85$): Left panels show LC for $R_{\epsilon} = 0.84$~k$\Omega$ ($\epsilon = 8.4$). Right panels show MNAD for $R_{\epsilon} = 1.4$~k$\Omega$ ($\epsilon = 14$). For other parameters see text. Left panel, $y$ axis: (a) 156 mv/div (c) 300 mv/div (e) 128 mv/div; $x$ axis: 780 $\mu$s/div. Right panel, $y$ axis: (a) 50 mv/div (c) 350 mv/div (e) 100 mv/div; $x$ axis: 10 ms/div.}
\end{figure}
\begin{subequations}
\label{evlv}
\begin{align}
\label{lxv}
10^5C\dot{V}_{xi} &= \frac{10^5}{R_\sigma}(V_{yi}-V_{xi})+\mathcal{F}_{\mathcal{C}},\\
\label{lyv}
10^5C\dot{V}_{yi} &= \frac{10^4}{R_r}(V_a-V_{zi})V_{xi}-V_{yi},\\
\label{lzv}
10^5C\dot{V}_{zi} &= V_{xi}V_{yi}-\frac{10^5}{R_b}V_{zi},\\
\label{lsv}
10^5C\dot{V}_{s} &= -\frac{10^5}{R_k}V_{s}-\frac{R_{\epsilon}}{100}\bigg(\frac{V_{x1}+V_{x2}}{2}\bigg).
\end{align}
\end{subequations}
Where $\mathcal{F}_{\mathcal{C}} = \frac{R_d}{100}(V_{xj}-V_{xi})+\frac{R_{\epsilon}}{100}V_s$ with $i,~j = 1,~2$ and $i \neq j$. Equation \eqref{evlv} is normalized with respect to $10^5C$ and thus now becomes equivalent to Eq.~\eqref{evlexpt} for the following normalized parameters: $\dot{u} = \frac{du}{d\tau}$, $\tau = \frac{t}{10^5C}$, $\epsilon = \frac{R_{\epsilon}}{100}$, $d = \frac{R_d}{100}$,  $k=\frac{10^5}{R_k}$, $\sigma = \frac{10^5}{R_{\sigma}}$, $\frac{r}{3} = \frac{10^4}{R_r}$, $b = \frac{10^5}{R_b}$, $V_a = 3$, $x_i=\frac{V_{xi}}{V_{sat}}$, $y_i=\frac{V_{yi}}{V_{sat}}$, $z_i=\frac{V_{zi}}{V_{sat}}$, and $s=\frac{V_s}{V_{sat}}$. Thus, the resistances $R_d$, $R_{\epsilon}$ and $R_k$ control the diffusive coupling strength ($d$), environmental coupling strength ($\epsilon$) and the damping factor of the environment ($k$), respectively. $V_{sat}$ is the op-amp saturation voltage. We choose  C=10 nF,  $R_r=1.07$~k$\Omega$ ($r=28$), $R_{\sigma}=10$~k$\Omega$ ($\sigma = 10$), $R_b=39$~k$\Omega$ ($b=2.6$), and $V_a=3$ volt. These particular choice of parameter values make the system represented by Eq.~\eqref{evlv} equivalent to that given by Eq.~\eqref{evlexpt} and keep the uncoupled Lorenz systems in the chaotic region.

At first, we take $R_{\epsilon} = 1.127$~k$\Omega$ ($\epsilon = 11.27$), $R_k = 100$~k$\Omega$ ($k=1$) and observe a continuous transition from the limit cycle to MMD for decreasing $R_d$ ($d$). In Figs. \ref{exptmmd}(a), (c), (e) using the experimental snapshots of the wave forms [taken using a digital storage oscilloscope (Agilent, DSO-X 2024A, 200MHz, 2 GSa/s)], we experimentally observed two distinct dynamical regions for two different $R_d$~$(d)$ values. (1) Limit cycle (LC): For $R_d = 4.55$~k$\Omega$ i.e. $d=45.5$ [Figs. \ref{exptmmd}(a), (c), (e) left panel]. (2) MMD: For $R_d = 145.6$~$\Omega$ i.e., $d=1.456$ [Figs.~\ref{exptmmd}(a), (c), (e) right panel]. We define this MMD state (see Sec.~\ref{sec2}) as the mixture of OD and AD state. The experimental real time traces clearly show this simultaneous occurrence of OD in $V_x$ [Fig.~\ref{exptmmd} (a) right panel] and $V_y$ [Fig. \ref{exptmmd}(c) right panel] and  AD (AD2) state in $V_z$ [Fig. \ref{exptmmd}(e) right panel]. The numerical time series plots (using the fourth-order Runge-Kutta method with 0.01 step size) for the equivalent parameter values are shown in Figs. \ref{exptmmd}(b),(d) and (f), which clearly shows the {\it qualitative} agreement between the experimental and numerical results. However, the slight mismatch between the experimental and numerical results may be due to the possible parameter mismatch and fluctuations that are inevitable in a real circuit.

Next, we set $R_d = 8.5$~k$\Omega$ ($d=85$), $R_k = 100$~k$\Omega$ ($k=1$) and observe a continuous transition from the limit cycle to MNAD for increasing $R_{\epsilon}$.  The experimental results are shown in Figs. \ref{exptmnad}(a), (c) and (e). Here the following observations are made (1) LC: For $R_{\epsilon} = 0.84$~k$\Omega$ ($\epsilon = 8.4$) [Figs.~\ref{exptmnad}(a), (c), (e) left panel]. (2) MNAD: For $R_{\epsilon} = 1.4$~k$\Omega$ ($\epsilon = 14$) [Fig. \ref{exptmnad}(a), (c), (e), right panel]. The numerical time series plots are demonstrated in Figs.~\ref{exptmnad}(b),(d),(f) for the equivalent parameter values. Here also, Fig. \ref{exptmnad} clearly shows that the MNAD state, as defined in  Sec.~\ref{sec2}, is a mixed state of  a bistable NAD and AD state. In the experimental study the bistability of the NAD state is found by a random parameter sweeping around $R_{\epsilon} = 1.4$~k$\Omega$, and the same is verified in numerical simulations using the proper initial conditions (not shown in Fig.~\ref{exptmnad}).

\section{CONCLUSION}
\label{sec:con}
\noindent We have reported a new cooperative dynamical state, namely the {\it mixed mode oscillation suppression state}, where different set of variables of a system of coupled oscillators show different types of oscillation suppression states under the same parametric condition. We identify two types of this state in coupled chaotic Lorenz oscillators: One is called the mixed mode death (MMD) state, where OD and AD occurs simultaneously to different set of variables, and the other one (called the MNAD state) is the mixed variable selective state of nontrivial bistable AD and a monostable AD. To show the generality of the results we consider two generic coupling schemes, namely direct-indirect coupling and mean-field coupling, which were studied earlier in the context of AD and OD. Also, we verify the results in the coupled Chen system, which is a Lorenz-like system. Through rigorous bifurcation analyses we find all the transition routes to these mixed   oscillation suppression states and map them in parameter space. We identify the underlying symmetry breaking that leads to the MMD and MNAD states. Finally, we report the first experimental observation of the MMD and MNAD state using an electronic circuit experiment.

The present study may have applications in many real systems, such as laser \cite{haken-lorenz} and geomagnetic \cite{geo} systems, whose models mimic the Lorenz system (under some proper transformations). Take for example of the laser system modeled by Maxwell-Bloch equation \cite{haken-lorenz} where the variables related to the electric-field and polarization mimic $x$ and $y$ variables, respectively, whereas the variable related to population inversion mimics the $z$ variable of the Lorenz system. Thus, we believe that the results of the present study can be extended to other ``Lorenz--like'' {\it physical systems} and may be useful in understanding of those systems.


\begin{acknowledgments}
T. B. acknowledges the financial support from SERB, Department of Science and Technology (DST), India [Project Grant No.: SB/FTP/PS-005/2013]. D. G. acknowledges DST, India, for providing support through the INSPIRE fellowship.
\end{acknowledgments}


\providecommand{\noopsort}[1]{}\providecommand{\singleletter}[1]{#1}%
\end{document}